\documentclass[a4paper,aps,prd,nofootinbib,onecolumn,notitlepage]{revtex4}

\RequirePackage[english]{babel}
\RequirePackage[latin1]{inputenc}
\RequirePackage[T1]{fontenc}
\RequirePackage{mathrsfs}
\RequirePackage{amsmath}
\RequirePackage{amssymb}
\RequirePackage{amsbsy}
\RequirePackage{bm}
\usepackage[lofdepth,lotdepth]{subfig}
\usepackage{graphicx}
\usepackage{multirow} 
\usepackage{caption}
\def\de#1/de#2{\frac{\partial {#1}}{\partial {#2}}}
\def\a{\alpha}
\def\b{\beta}

\def\m{\mu}
\def\n{\nu}

\def\d{\delta}

\newcommand{\ba}{\begin{eqnarray}}
\newcommand{\ea}{\end{eqnarray}}
\newcommand{\be}{\begin{equation}}
\newcommand{\ee}{\end{equation}}

\begin{document}


\title{Quark stars with isotropic matter in Ho\v rava gravity and Einstein-{\ae}ther theory}

\author{Grigorios Panotopoulos$^1$, Daniele Vernieri$^2$, and Ilidio Lopes$^1$}

\affiliation{
\smallskip
$^1$Centro de Astrof{\'i}sica e Gravita\c c\~ao-CENTRA, Departamento de F{\'i}sica,
Instituto Superior T{\'e}cnico-IST, Universidade de Lisboa-UL,
Avenida Rovisco Pais 1, 1049-001, Lisboa, Portugal
\smallskip \\
$^2$Instituto de Astrof\'isica e Ci\^encias do Espa\c{c}o, 
Faculdade de Ci\^encias da Universidade de Lisboa, Campo Grande, 
PT1749-016 Lisboa, Portugal
}

\date{\today}


\begin{abstract}
We study non-rotating and isotropic strange quark stars in Lorentz-violating theories of gravity, and in particular in Ho\v rava gravity and Einstein-{\ae}ther theory. For quark matter we adopt both linear and non-linear equations of state, corresponding to the MIT bag model and color flavor locked state, respectively. The new structure equations describing hydrostatic equilibrium generalize the usual Tolman-Oppenheimer-Volkoff (TOV) equations of Einstein's general relativity. A dimensionless parameter $\nu$ measures the deviation from the standard TOV equations, which are recovered in the limit $\nu \rightarrow 0$. We compute the mass, the radius as well as the compactness of the stars, and we show graphically the impact of the parameter $\nu$ on the mass-to-radius profiles for different equations of state describing quark matter. The energy conditions and stability criteria are also considered, and they are all found to be fulfilled.
\end{abstract}

\maketitle


\section{Introduction} 

In 2009 Ho\v rava gravity~\cite{Horava:2009uw,Blas:2009qj} was proposed as a new candidate theory for quantum gravity which explicitly breaks Lorentz invariance at any energy scale by introducing a preferred foliation of spacetime. Since then a lot of work has been done to prove, very successfully, its renormalizability by means of both power-counting arguments~\cite{Visser:2009fg,Sotiriou:2009gy,Sotiriou:2009bx,Vernieri:2011aa,Vernieri:2012ms,Vernieri:2015uma} and quantum field theory approaches~\cite{DOdorico:2014tyh,Barvinsky:2015kil,Barvinsky:2017zlx,Barvinsky:2019rwn}. 
Moreover, a big effort has also been made in order to unveil its phenomenological implications, e.g. concerning late-time cosmology~\cite{Audren:2014hza,Frusciante:2015maa}, black holes~\cite{Blas:2011ni,Barausse:2011pu,Berglund:2012bu,Barausse:2012qh,Wang:2012nv,Barausse:2013nwa,Sotiriou:2014gna}, binary systems~\cite{Yagi:2013qpa,Yagi:2013ava,Ramos:2018oku}, and anisotropic interior solutions~\cite{Vernieri:2017dvi,Vernieri:2018sxd,Vernieri:2019vlh}. After the multiple detections of gravitational waves by the LIGO-VIRGO Collaboration, and in particular the first merger observed from a binary of neutron stars~\cite{Monitor:2017mdv}, a new era for gravitational-wave astronomy just got started. Very interestingly, Ho\v rava gravity passes with flying colors all the theoretical and observational constraints which are available to date~\cite{Gumrukcuoglu:2017ijh}. Notice that if one takes the low-energy limit of Ho\v rava gravity and writes its action in a covariant form, the latter becomes equivalent to Einstein-\ae ther theory~\cite{Jacobson:2000xp} once the \ae ther vector is restricted to be hypersurface-orthogonal at the level of the action~\cite{Jacobson:2010mx}. In spherical symmetry it can be shown that the two theories share the same solutions~\cite{Blas:2010hb}.

Therefore, it has become even more urgent to study the implications and predictions of viable alternative gravity theories at astrophysical scales, in order to explore non-standard scenarios and the possible signatures of deviations from general relativity (GR) to be observed in the forthcoming detections. For all of these reasons in this work we will investigate some astrophysical implications of the theory. In this respect, compact objects~\cite{textbook,review1,review2} such as neutron stars and white dwarfs, are relativistic stars of astrophysical and astronomical interest, which are characterized by ultra dense matter densities and strong gravitational fields, and thus they serve as ideal cosmic laboratories to study and test non-standard physics as well as non-conventional theories of gravity.

A new class of compact objects, that may be an alternative to neutron stars, are some as of today hypothetical objects which are supposed to be made of quark matter, and for that reason they are called strange quark stars~\cite{SS1,SS2,SS3,SS4,SS5,SS6}. Quark matter is by assumption absolutely stable~\cite{witten,farhi}, and so it could be the true ground state of hadrons. That property makes them a plausible explanation of some puzzling super-luminous supernovae~\cite{SL1,SL2}, which occur in about one out of every 1000 supernovae explosions, and which are more than 100 times more luminous than regular supernovae.

The plan of our work is the following. In Sec.~\ref{Sec1} we briefly review the basic ingredients of Ho\v rava gravity and its connection to Einstein-{\ae}ther theory, while in Sec.~\ref{Sec2} we present the field equations as well as the structure equations describing the hydrostatic equilibrium of spherically symmetric relativistic stars with isotropic matter. In Sec.~\ref{Sec3} we obtain and discuss our numerical results for quark stars. Finally we finish our work with some conclusions in Sec.~\ref{Sec4}. We adopt the mostly negative metric signature $+,-,-,-$, and we work in units in which the speed of light in vacuum $c$ as well as the reduced Planck constant $\hbar$ are set equal to unity, $\hbar = 1 = c$. In those units all dimensionful quantities are measured in GeV $=1000$~MeV, and we make use of the conversion rules
$1$~m $= 5.068 \times 10^{15}$~GeV$^{-1}$ and $1$~kg $= 5.610 \times 10^{26}$~GeV~\cite{guth}.

\section{Ho\v rava gravity and Einstein-\ae ther theory}
\label{Sec1}

The action of Ho\v rava gravity~\cite{Horava:2009uw,Blas:2009qj} can be written in the preferred foliation as
\be \label{horava}
\mathcal{S}_{H}=\frac{1}{16\pi G_H}\int{dT d^3x\sqrt{-g}\left(K_{ij}K^{ij}-\lambda K^2 +\xi \mathcal{R}+\eta a_i a^i+\frac{L_4}{M_\ast^2}+\frac{L_6}{M_\ast^4}\right)}+S_m[g_{\mu\nu},\psi]\,,
\ee 
where $G_H$ is the effective gravitational constant; $g$ is the determinant of the metric $g_{\mu\nu}$; $\mathcal{R}$ is the Ricci scalar of the three-dimensional constant-$T$ hypersurfaces; $K_{ij}$ is the extrinsic curvature and $K$ is its trace; and $a_i=\partial_i \mbox{ln} N$, where $N$ is the lapse function and $S_m$ is the matter action where $\psi$ collectively denotes the matter fields. The constant couplings $\left\{\lambda,\xi,\eta\right\}$ are dimensionless, and GR is identically recovered when they take the values $\left\{1,1,0\right\}$, respectively. Moreover, $L_4$ and $L_6$ denote the fourth-order and sixth-order operators respectively, while $M_\ast$ is the characteristic mass scale which suppresses them at low-energy. 

In the following, we consider the covariantized version of the low-energy limit of Ho\v rava gravity, named the {\it khronometric} model, that is obtained by keeping only the operators up to second-order derivatives, which amounts to discarding $L_4$ and $L_6$ which instead contain the higher-order operators. 

In order to write the action covariantly, let us first take the action of Einstein-\ae ther theory~\cite{Jacobson:2000xp}:
\be
\mathcal{S}_{\mbox{\scriptsize\ae}}=\frac{1}{16\pi G_{\mbox{\scriptsize\ae}}}\int{d^4x\sqrt{-g}\left(-R+L_{\mbox{\scriptsize\ae}}\right)}+S_m[g_{\mu\nu},\psi]\,, \label{aetheraction}
\ee
where $G_{\mbox{\scriptsize\ae}}$ is the ``bare'' gravitational constant; $R$ is the four-dimensional Ricci scalar; $u^a$ is a timelike vector field of unit norm, {\it i.e.}, $g_{\mu\nu}u^\mu u^\nu=1$, from now on referred to as the ``\ae ther''; and 
\be
L_{\mbox{\scriptsize\ae}}=-M^{\a\b}{}_{\m\n} \nabla_\a u^\m \nabla_\b u^\n\,,
\ee
with $M^{\a\b}{}_{\m\n}$ defined as
\be 
M^{\a\b}{}_{\m\n} = c_1 g^{\a\b}g_{\m\n}+c_2\d^{\a}_{\m}\d^{\b}_{\n}+c_3 \d^{\a}_{\n}\d^{\b}_{\m}+c_4 u^\a u^\b g_{\m\n}\,,
\ee
where $c_i$'s are dimensionless coupling constants.

Once the \ae ther vector is taken to be hypersurface-orthogonal at the level of the action, that is
\be
u_\alpha=\frac{\partial_\alpha T}{\sqrt{g^{\mu\nu}\partial_\mu T \partial_\nu T}}\,,
\ee
where the preferred time $T$ is a scalar field (the {\it khronon}) which defines the preferred foliation, then the two actions in Eqs.~\eqref{horava} and~\eqref{aetheraction} become equivalent if the parameters of the two theories are mapped into each other as~\cite{Jacobson:2010mx}
\be
\label{eqn:corresp}
\frac{G_H}{G_{\scriptsize\mbox{\ae}}}=\xi = \frac{1}{1-c_{13}}\,, \hspace{2em} \frac{\lambda}{\xi} = 1 + c_2\,, \hspace{2em} \frac{\eta}{\xi} = c_{14}\,,
\ee
where $c_{ij} = c_i+c_j$. 
Moreover $G_{\mbox{\footnotesize \ae}}=G_N \left(1-\eta/2\xi \right)$, where $G_N$ is the Newton's constant, which is needed to recover the Newtonian limit~\cite{Carroll:2004ai,Blas:2009qj}. 
In what follows, we will thus consider the covariant formulation of the low-energy limit of Ho\v rava gravity. 

The variation of the action in Eq.~\eqref{aetheraction} with respect to $g^{\a\b}$ and $T$ yields, respectively~\cite{Ramos:2018oku},
\ba 
\label{eqhorava}
&&G_{\a\b} - T^{\mbox{\scriptsize\ae}}_{\a\b}=8\pi G_{\mbox{\scriptsize\ae}} T^{m}_{\a\b}\,,\\
\label{hleq}
&&\partial_\mu \left(\frac{1}{\sqrt{\nabla^\alpha T \nabla_\alpha T}}\sqrt{-g} \mbox{\AE}^\mu \right)=0\,,
\ea
where
$G_{\a\b}=R_{\a\b}-R g_{\a\b}/2$ is the Einstein tensor,
\ba \label{Tae}
T^{\mbox{\scriptsize\ae}}_{\a\b}&=&\nabla_\m\left(J^{\phantom{(\a}\m}_{(\a}u_{\b)}-J^\m_{\phantom{\m}(\a}u_{\b)}-J_{(\a\b)}u^\m\right)+c_1\,\left[ (\nabla_\m u_\a)(\nabla^\m u_\b)-(\nabla_\a u_\m)(\nabla_\b u^\m) \right]\nonumber\\
&&+\left[ u_\n(\nabla_\m J^{\m\n})-c_4 \dot{u}^2 \right] u_\a u_\b
+c_4 \dot{u}_\a \dot{u}_\b-\frac{1}{2} L_{\mbox{\scriptsize\ae}} g_{\a\b} + 2 \mbox{\AE}_{(\a}u_{\b)}
\ea
is the khronon stress-energy tensor, 
\be
J^\a_{\phantom{a}\m}=M^{\a\b}_{\phantom{ab}\m\n} \nabla_\b u^\n\,,
\qquad\dot{u}_\n=u^\m\nabla_\m u_\n\,,\qquad \mbox{\AE}_\mu = \left(\nabla_\a J^{\a\n}-c_4\dot{u}_\a\nabla^\n u^\a\right) \left(g_{\mu\nu}-u_\m u_\n\right)\,,
\ee
and $T^{m}_{\a\b}$ is the matter stress-energy tensor, defined as
\be
T^{m}_{\a\b} = \frac{2}{\sqrt{-g}}\frac{\delta S_m}{\delta g^{\a\b}}\,.
\ee

\section{Field Equations}
\label{Sec2}

The most general static and spherically symmetric metric, in Schwarzschild coordinates, 
can be written as
\be
ds^2 = e^{A(r)} dt^2 - B(r) dr^2 - r^2\,\big(d\theta^2+\sin^2\theta d\phi^2\big)\,. \label{Eq0}
\ee
In addition, let us consider an interior spacetime filled by an isotropic fluid whose stress-energy tensor is 
\be
T^{m}_{\alpha\beta}=\left(\rho + p\right) v_\alpha v_\beta - p g_{\alpha\beta}\,,
\ee
where $\rho$ is the energy density and $p$ is the pressure of the fluid, and its 4-velocity $v^\alpha$ is given by
\be
v^\alpha =\biggl(e^{-A(r)/2},0,0,0\biggr)\,.
\ee
The \ae ther vector field, which is by definition a unit timelike vector, in spherical symmetry is always hypersurface-orthogonal and takes the following general form: 
\be
u^\alpha =\biggl(F(r),\sqrt{\frac{e^{A(r)} F(r)^2-1}{B(r)}},0,0\biggr)\,. \label{aether}
\ee
However, in the following we will consider the case of a static \ae ther which is aligned with the interior matter fluid 4-velocity $v^\alpha$, {\it i.e.} when $F(r)=e^{-A(r)/2}$, which leads to
\be
u^\alpha =\biggl(e^{-A(r)/2},0,0,0\biggr)\,.
\ee
The field equations that we have to consider are the modified Einstein equations (0-0), (1-1) and (2-2) in Eq.~\eqref{eqhorava}, which can be written respectively as: 
\be
- \nu \left[4 r^2 A''(r)-\frac{2 r^2 A'(r) B'(r)}{B(r)}+ r^2 A'(r)^2+8 r A'(r)\right]+\frac{r B'(r)}{B(r)}+B(r)-1=8 \pi G_{\mbox{\scriptsize\ae}} r^2  B(r) \rho (r)\,, \label{eq1}
\ee
\be
\nu  r^2 A'(r)^2+r A'(r)-B(r)+1=8 \pi G_{\mbox{\scriptsize\ae}} r^2 B(r) p(r)\,, \label{eq2}
\ee
\be
\frac{1}{2} r^2 A''(r)-\frac{r^2 A'(r) B'(r)}{4 B(r)}-\nu  r^2 A'(r)^2+\frac{1}{4} r^2 A'(r)^2+\frac{1}{2} r A'(r)-\frac{r B'(r)}{2 B(r)}=8 \pi G_{\mbox{\scriptsize\ae}} r^2 B(r) p(r)\,, \label{eq3}
\ee
where $\nu = \frac{\eta}{8\xi}$. Furthermore, Eq.~\eqref{hleq} is identically satisfied. In the following we shall set $8 \pi G_{\mbox{\scriptsize\ae}} = 1$. \\
Moreover, the conservation equation for the stress-energy tensor is
\be
p'(r)+\frac{1}{2}A'(r)\left[\rho(r)+p(r)\right]=0\,. \label{conserv}
\ee
Notice that the standard Tolman-Oppenheimmer-Volkoff (TOV) equations~\cite{Tolman:1939jz,OV,Carloni:2017rpu,Carloni:2017bck} which hold in GR~\cite{GR} are recovered by setting $\nu=0$ in the equations above.

Nevertheless, among the above 4 equations only 3 are independent. Furthermore, since spherically symmetric solutions in Ho\v rava gravity are identical to those of Einstein-\ae ther theory, all of our conclusions will hold for both theories~\cite{Blas:2010hb}. In order to derive the modified TOV equations that will be used for the numerical integration, one can obtain $A'(r)$ from Eq.~\eqref{eq2}: 
\be
A'(r)=-\frac{r-r \sqrt{4 \nu  r^2 B(r) p(r)+4 \nu  B(r)-4 \nu +1}}{2 \nu  r^2}\,,
\ee
which corresponds to the only branch which admits the proper GR limit when $\nu\rightarrow 0$.
Then, one has to substitute the latter in Eq.~\eqref{eq1} and Eq.~\eqref{conserv}, that for brevity we do not show here. However, one can immediately notice the difference with respect to GR, since here the parameter $\nu$ (which in GR is identically zero) enters non-linearly in the resulting equations.

\section{Properties of strange quark stars: Numerical treatment}
\label{Sec3}

In the present section we investigate the properties of quark stars in the Lorentz-violating theories of gravity at hand. We integrate the structure equations numerically, and then we present and discuss our main results.

\subsection{Vacuum Solution}

To be able to match the solutions at the surface of the stars in order to compute the mass of the objects we need to know the exterior (vacuum) solution first. To that end we set the stress-energy tensor (pressure and energy density of the fluid) to zero.

In order to calculate the total gravitational mass $M$ of the star appearing in the Newtonian potential, we will make use of the general vacuum solution found in Ref.~\cite{Eling:2006df} and summarized below:
\be
e^A=\left(\frac{1-Y/Y_-}{1-Y/Y_+}\right)^{\frac{-Y_+}{2+Y_+}}\,,
\ee
\be 
B=\nu (Y-Y_-)(Y-Y_+)\,,
\ee
\be 
\frac{r_{\rm
min}}{r}=\left(\frac{Y}{Y-Y_-}\right)\left(\frac{Y-Y_-}{Y-Y_+}\right)^{\frac{1}{2+Y_+}}\,, 
\ee
where $Y=rA'$, $Y_\pm=(-1\pm\sqrt{1-4\nu})/(2\nu)$, and $r_{\rm min}$ is an integration constant which is related to the gravitational radius $r_g=2 G_N M$ by
\be
r_{\rm min}/r_g=(-Y_+)^{-1}(-1-Y_+)^{(1+Y_+)/(2+Y_+)}\,.
\ee
The above solution agrees with the Schwarzschild solution~\cite{SBH} of GR to leading order in $1/r$.

\subsection{Equation-of-state}

Before we continue to integrate the structure equations, we must specify the sources first.
Quark matter inside the stars is described by the MIT bag model~\cite{bagmodel1,bagmodel2}, where in the simplest version there is a linear analytic function relating the energy density to the pressure of the fluid, that is
\begin{equation}
p = k (\rho - \rho_0)\,,
\end{equation}
where $k$ is a dimensionless numerical factor, while $\rho_0$ is the surface energy density.
The MIT bag model is characterized by 3 parameters, namely i) the QCD coupling constant, $\alpha_c$, ii) the mass of the strange (s) quark, $m_s$, and iii) the bag constant, $B_0$. In this work we will consider the following 3 models~\cite{basic}:

\begin{itemize}
\item The extreme model SQSB40 where $m_s=100$~MeV, $\alpha_c=0.6$ and $B_0=40$~MeV~fm$^{-3}$\,.
In this model $k=0.324$ and $\rho_0=3.0563 \times 10^{14}$~g~cm$^{-3}$\,.
\item The standard model SQSB56 where $m_s=200$~MeV, $\alpha_c=0.2$ and $B_0=56$~MeV~fm$^{-3}$\,.
In this model $k=0.301$ and $\rho_0=4.4997 \times 10^{14}$~g~cm$^{-3}$\,.
\item The simplified model SQSB60 where $m_s=0=\alpha_c$ and $B_0=60$~MeV~fm$^{-3}$\,.
In this model $k=1/3$ and $\rho_0=4.2785 \times 10^{14}$~g~cm$^{-3}$\,.
\end{itemize}

Furthermore, at asymptotically large densities color superconductivity effects~\cite{wilczek1,wilczek2} become important. Quark matter is in the color flavor 
locked (CFL) state~\cite{wilczek3,wilczek4}, in which quarks form Cooper pairs of different color and flavor, and where all quarks have the same Fermi momentum and electrons cannot be present. That quark state is described by a slightly more complicated equation of state (EoS), although still an analytic function, and it is given by the following non-linear relation~\cite{CFL1,CFL2,CFL3}:
\begin{equation}
\rho = 3 p + 4 B_0 - \frac{9 \gamma \mu^2}{\pi^2}\,, 
\end{equation}
where $\gamma$ and $\mu^2$ are given by
\begin{equation}
\gamma = \frac{2 \Delta^2}{3} - \frac{m_s^2}{6}\,,
\end{equation}
and
\begin{equation}
\mu^2 = -3 \gamma + \left (9 \gamma^2 + \frac{4}{3} \pi^2 (B_0 + p) \right )^{1/2}\,,
\end{equation}
with $\Delta$ being the non-vanishing energy gap.

In the CFL state there are 19 viable models, but here we shall consider two, namely CFL4 and CFL10, characterized by the following parameters (see table I in Ref.~\cite{CFL3}):
\begin{eqnarray}
\Delta & = & 100~\text{MeV}\,, \\
m_s & = & 150~\text{MeV}\,, \\
B_0 & = & 60~\text{MeV}~\text{fm}^{-3}\,,
\end{eqnarray}
for CFL4, and
\begin{eqnarray}
\Delta & = & 150~\text{MeV}\,, \\
m_s & = & 150~\text{MeV}\,, \\
B_0 & = & 80~\text{MeV}~\text{fm}^{-3}\,,
\end{eqnarray}
for CFL10.

\subsection{Numerical solution of structure equations}

The radius of the stars, $R$, is determined from the requirement that $p(r=R)=0$, while the mass of the stars, $M$, is computed numerically using the vacuum solution presented before, and requiring that $B_{int}(r=R)=B_{ext}(r=R)$. Finally, the compactness of the object is computed by $C=G_N M/R$.

Our main numerical results are summarized in the figures below. At this point it should be mentioned that the 3 observed super-massive pulsars J1614-2230 ($(1.928 \pm 0.017)~M_{\odot}$)~\cite{shapiro,fonseca}, J0348+0432 ($(2.01 \pm 0.04)~M_{\odot}$)~\cite{antoniadis} and J0740+6620 ($(2.14_{-0.18}^{+0.20})~M_{\odot}$)~\cite{cromartie},
with masses $M \sim 2 M_{\odot}$ have put stringent constraints on compact object modelling, since any EoS that does not cross the 2 solar mass strip must be ruled out. Furthermore, the recent observation of the highly massive pulsar in the binary J2215+5135 ($(2.27 _{-0.15}^{+0.17})~M_{\odot}$)~\cite{Linares:2018ppq} has put an even more tight constraint to be satisfied. However, as it will be shown in the following, our results also depend on the specific EoS that is used, as it affects the predicted mass profiles. For this reason, we will not aim to show that our models are able to pass all the existing tests, but just to study the general features of quark stars in the framework of Ho\v rava gravity and Einstein-{\ae}ther theory and the overall effect of the modifications that they induce on the resulting mass profiles.
Moreover, we should also keep in mind that electrically charged stars or compact objects with anisotropic matter can have higher masses compared to their isotropic neutral counterparts, see e.g. Refs.~\cite{PR1,PR2} and references therein.

\begin{figure}[ht!]
\centering
\includegraphics[width=0.49\textwidth]{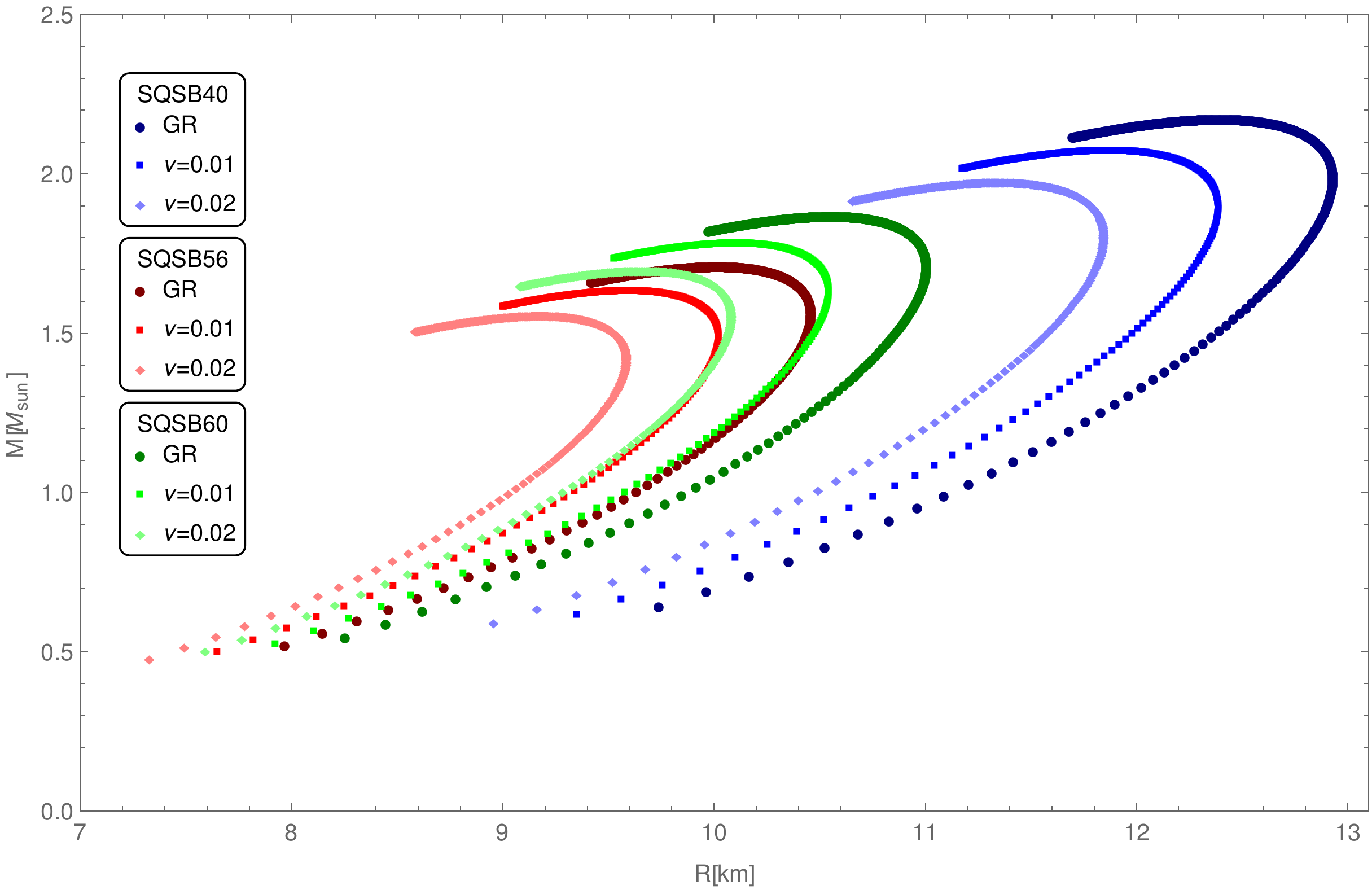} \ 
\includegraphics[width=0.49\textwidth]{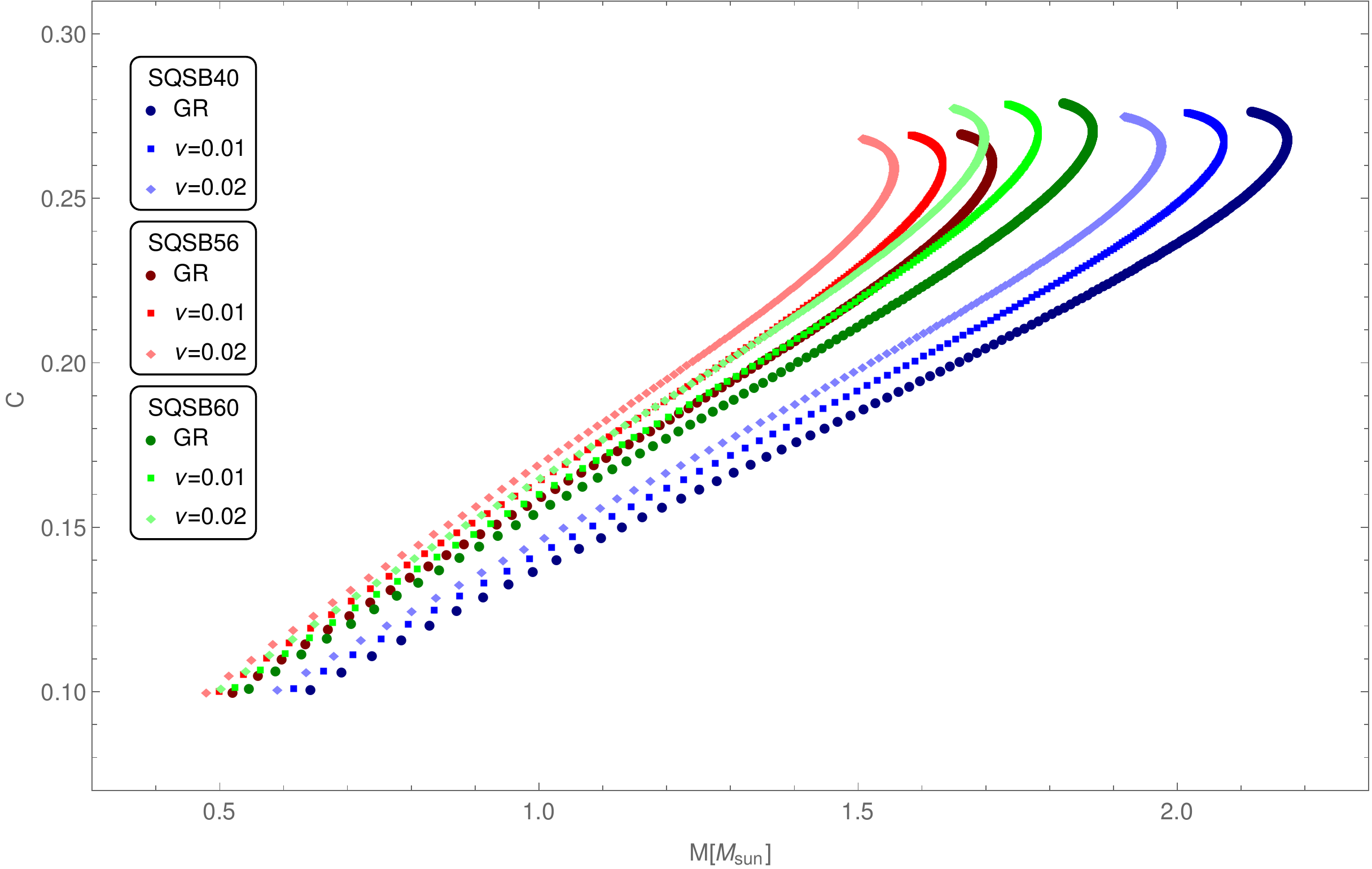}
\caption{
Properties of strange quark stars with linear EoS. We have considered 3 MIT bag models, SQSB40 (blue), SQSB56 (red) and SQSB60 (green) 
and $\nu=0.01, 0.02$. The curves corresponding to GR ($\nu=0$) are also shown for comparison reasons.
{\bf Left panel:} 
$M_R$ profiles (mass in solar masses and radius in km). 
{\bf Right panel:} 
Compactness $C=G_N M/R$ vs mass (in solar masses).
}
\label{fig:1}
\end{figure}

\begin{figure}[ht!]
\centering
\includegraphics[width=0.49\textwidth]{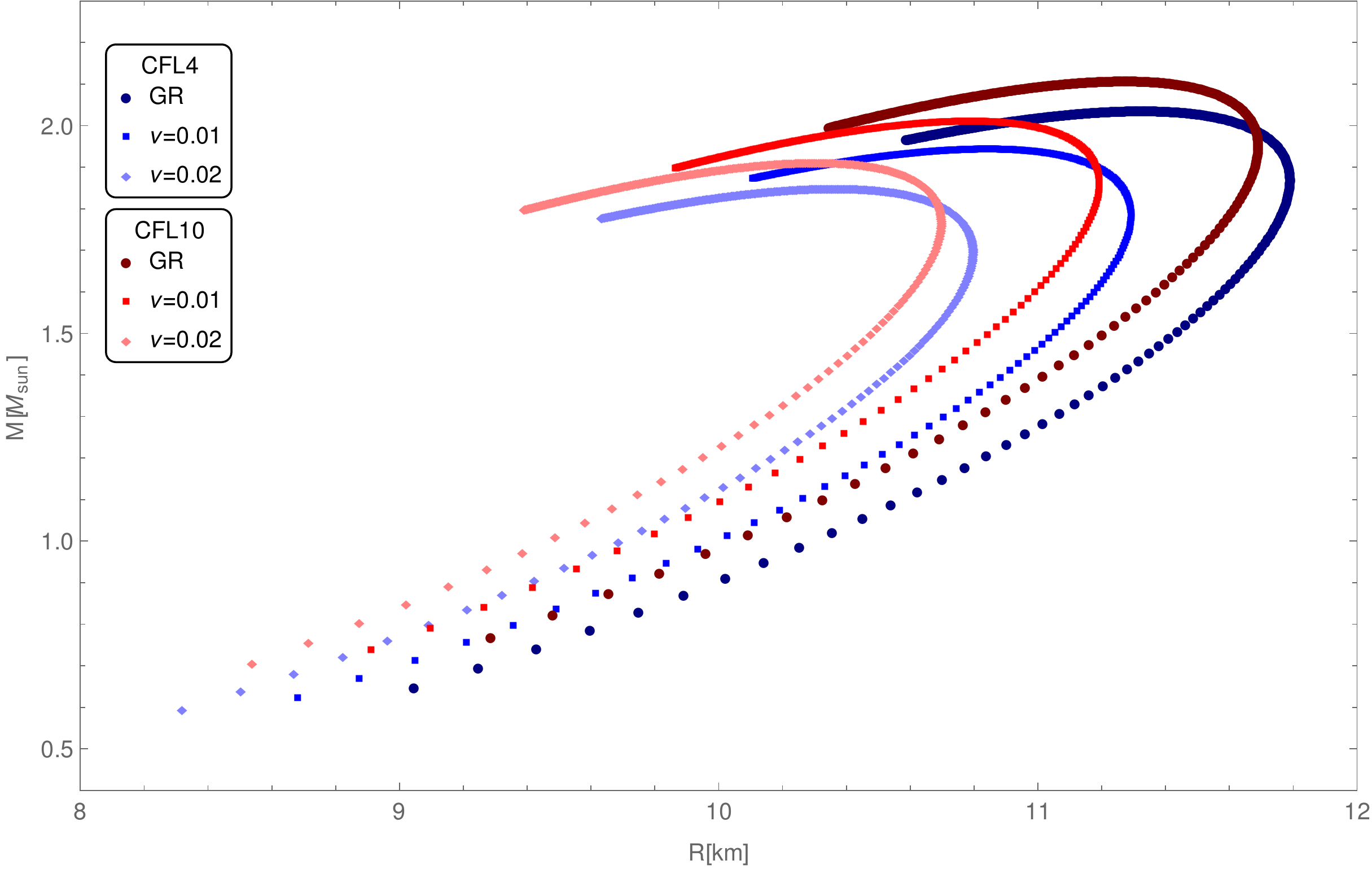} \ 
\includegraphics[width=0.49\textwidth]{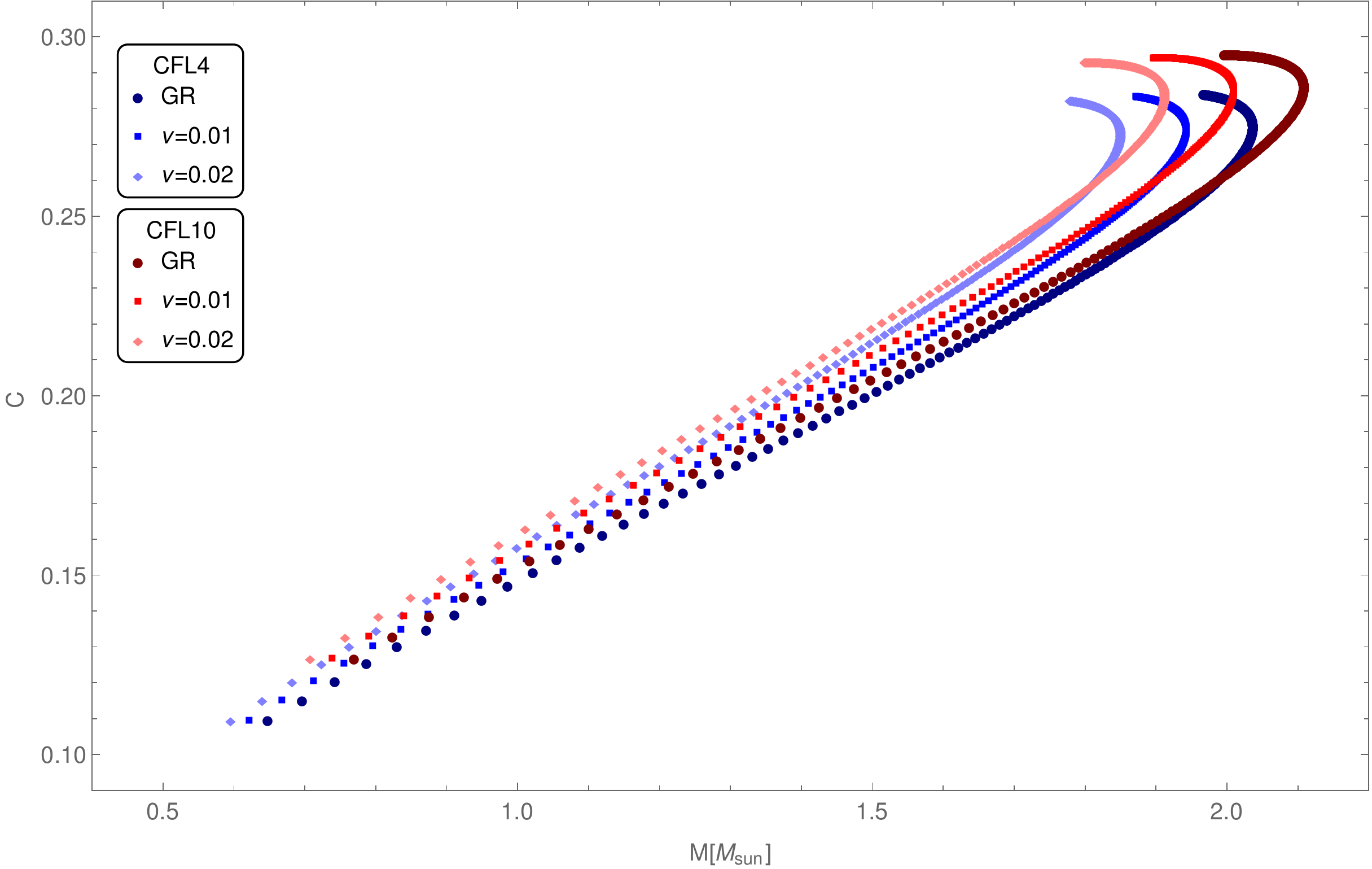}
\caption{
Properties of strange quark stars with a non-linear EoS. We have considered 2 CFL models, CFL4 (blue) and CFL10 (red) (see text) and $\nu=0.01, 0.02$. 
The curves corresponding to GR ($\nu=0$) are also shown for comparison reasons.
{\bf Left panel:} 
$M_R$ profiles (mass in solar masses and radius in km). 
{\bf Right panel:} 
Compactness $C=G_N M/R$ vs mass (in solar masses).
}
\label{fig:2}
\end{figure}

\begin{figure}[ht!]
\centering
\includegraphics[width=0.55\textwidth]{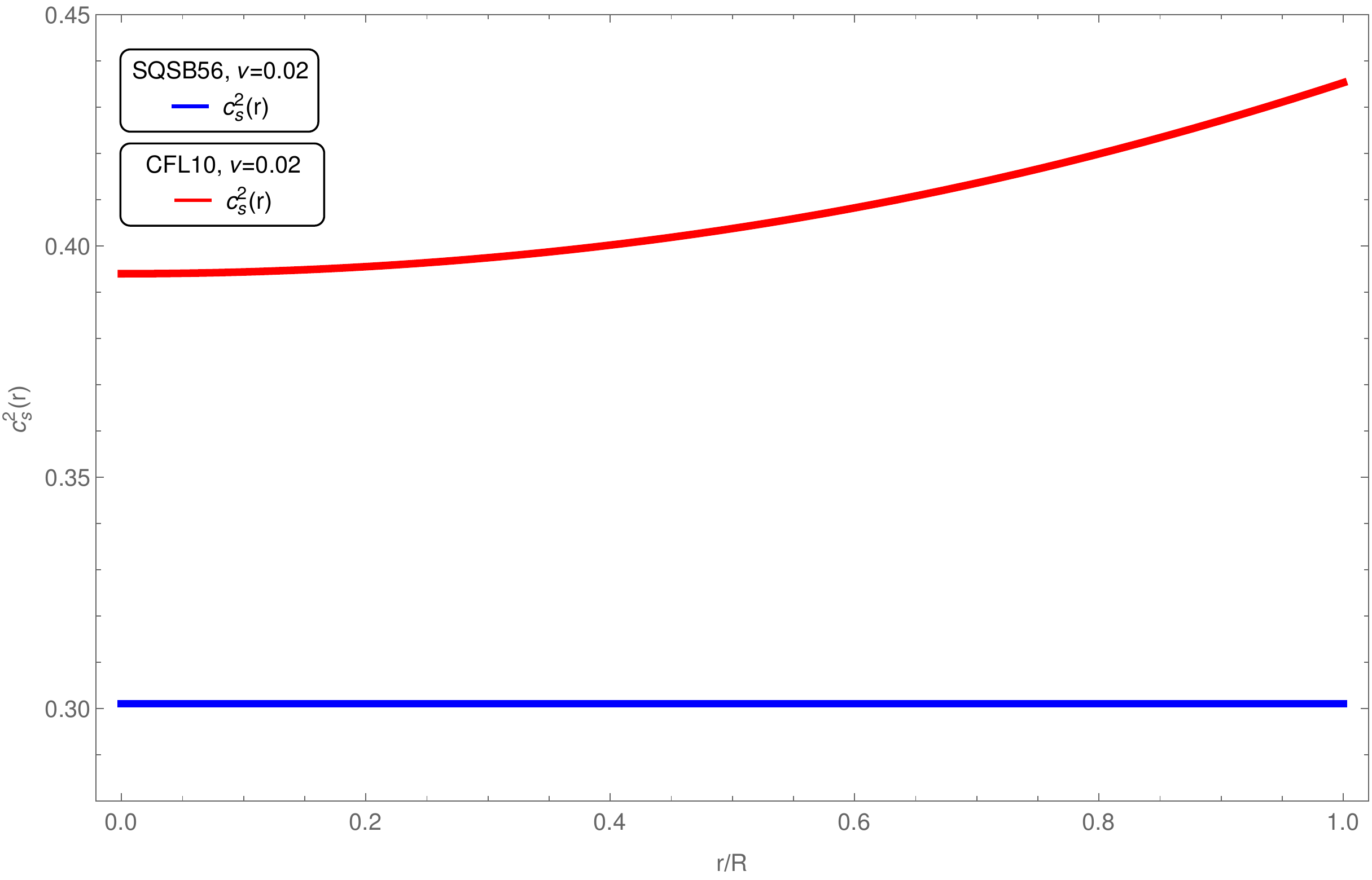} 
\caption{
Speed of sound, $c_s^2 = dp/d \rho$, vs normalized radial coordinate $r/R$ for $\nu=0.02$ and for the models SQSB56 (blue) and CFL10 (red).
}
\label{fig:3}
\end{figure}

In Fig.~\ref{fig:1} we show the mass-to-radius ($M_R$) relations (left panel) and the compactness (right panel) of the quark stars with a linear EoS. Similarly, Fig.~\ref{fig:2} shows the same properties of the stars with a non-linear EoS. We see that the $M_R$ profiles are shifted downwards as the parameter $\nu$ increases. Therefore, if the highest star mass that a given EoS can support is lower than the 2 solar mass limit in GR, it will become even worse in Ho\v rava gravity and Einstein-{\ae}ther theory. The speed of sound, defined by $c_s^2 \equiv dp/d \rho$, is shown in Fig.~\ref{fig:3} for the models SQSB56 and CFL10 for $\nu=0.02$, while for GR the speed of sound for the 19 CFL viable models can be seen in Fig.~2 (panel (b)) of Ref.~\cite{CFL3}. Clearly, throughout the object the speed of sound takes values in the range $0 < c_s^2 < 1$, as it should, and therefore causality is not violated. Moreover for each model we have checked that the 2 curves corresponding to $\nu=0.01, 0.02$ are indistinguishable. The latter is obvious when the EoS is linear since in that case the sound speed is just a constant, $c_s^2 = k$.

\begin{figure}[ht!]
\centering
\includegraphics[width=0.7\textwidth]{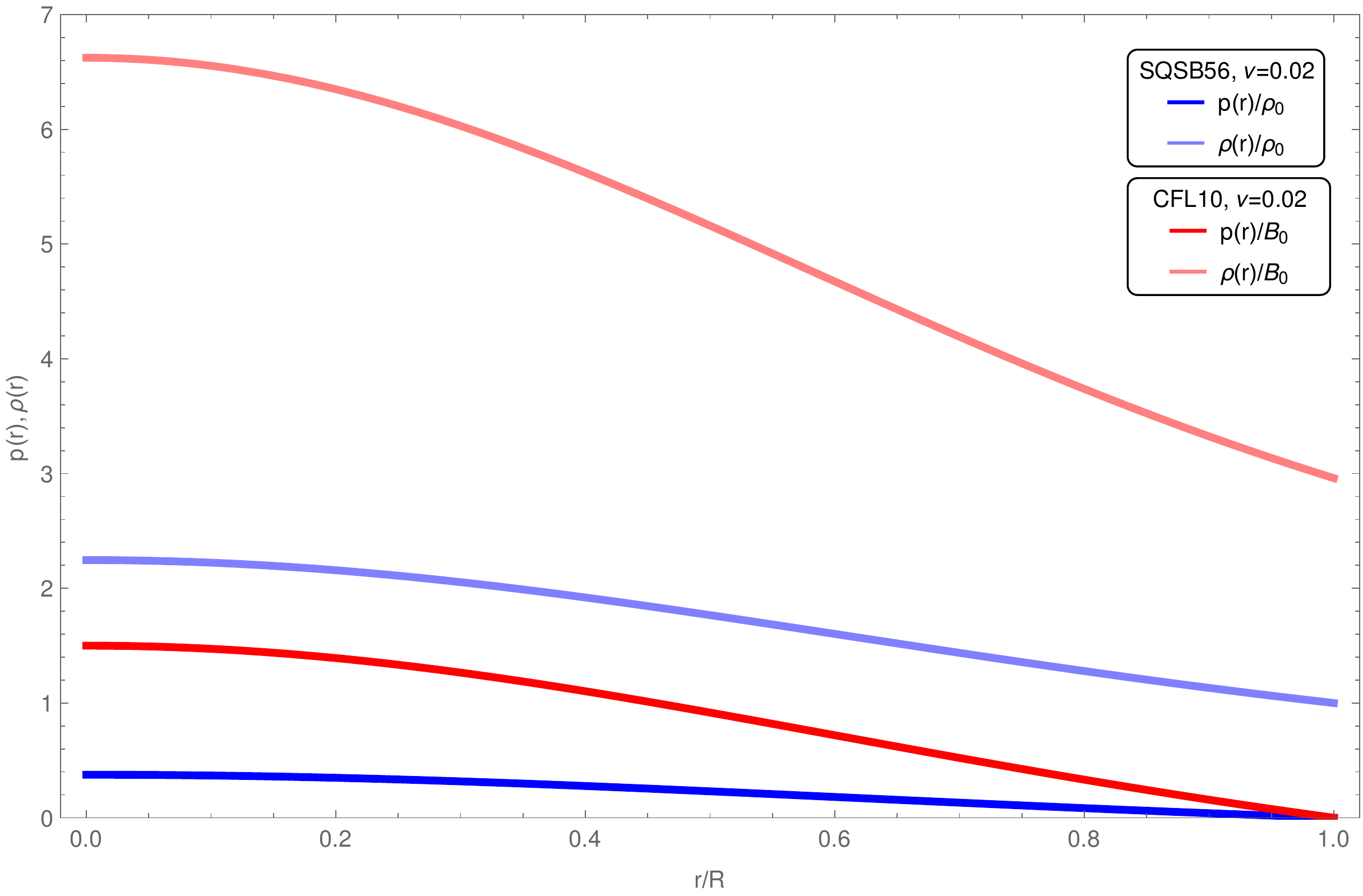}
\caption{
Normalized pressure and energy density vs normalized radial coordinate $r/R$ for $\nu=0.02$ for the models SQSB56 (blue) and CFL10 (red).}
\label{fig:4}
\end{figure}

The solutions obtained here should be able to describe realistic astrophysical configurations. Therefore, as a final check we investigate if the i) energy conditions and ii) stability criteria are fulfilled or not. Concerning the energy conditions, we require that~\cite{Ref_Extra_1,Ref_Extra_2,Ref_Extra_3,ultimo,PR3}:
\begin{equation}
\mbox{WEC:} \,\,\, \rho \geq 0\,, \,\,\, \rho + p \geq 0\,,
\end{equation}
\begin{equation}
\mbox{NEC:} \,\,\, \rho + p  \geq  0\,,
\end{equation}
\begin{equation}
\mbox{DEC:} \,\,\, \rho \geq \lvert p \rvert\,,
\end{equation}
\begin{equation}
\mbox{SEC:} \,\,\, \rho + p  \geq  0\,, \,\,\, \rho + 3p \geq 0\,.
\end{equation} 
In Fig.~\ref{fig:4} we show as an example the normalized pressure, and normalized energy density, versus the normalized radial coordinate $r/R$ for $\nu=0.02$, $p_c(0)=1.5~\rho_0/4$ and $p_c(0)=1.5~B_0$ for the models SQSB56 and CFL10, respectively. Let us mention that within each model the 2 curves corresponding to $\nu=0.01, 0.02$ cannot be told apart. Regarding all the other models we have obtained qualitatively very similar curves that are not shown here. Clearly, the energy conditions are fulfilled throughout the star, and we thus conclude that the solutions obtained in the present work are realistic solutions, which are able to describe realistic astrophysical configurations.

\begin{figure}[ht!]
\centering
\includegraphics[width=0.7\textwidth]{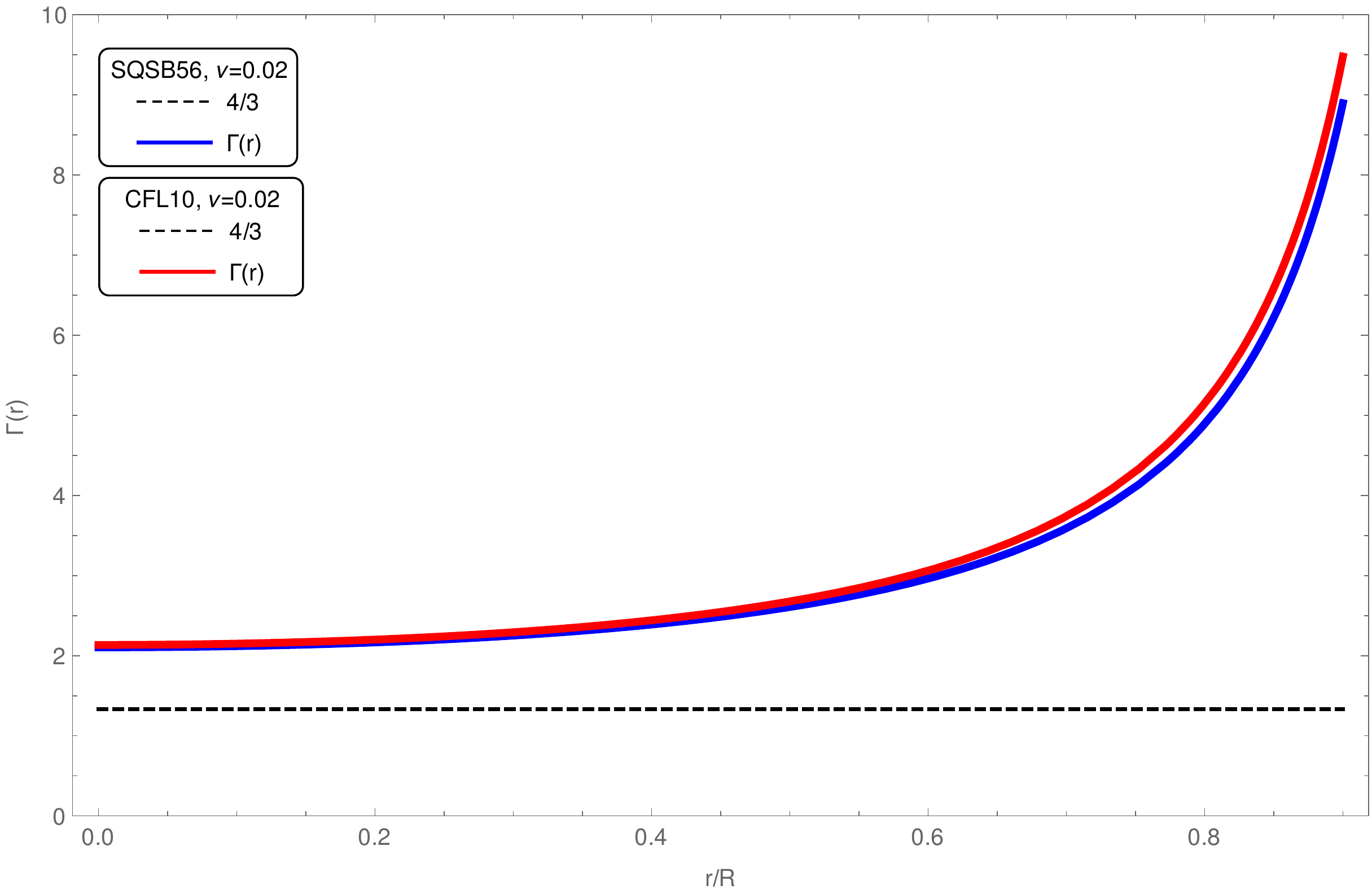}
\caption{
Adiabatic index $\Gamma$ vs normalized radial coordinate $r/R$ for $\nu=0.02$ for the models SQSB56 (blue)
and CFL10 (red). The dashed horizontal line represents the Newtonian 
bound corresponding to $4/3$.
}
\label{fig:5}
\end{figure}

\begin{figure}[ht!]
\centering
\includegraphics[width=0.49\textwidth]{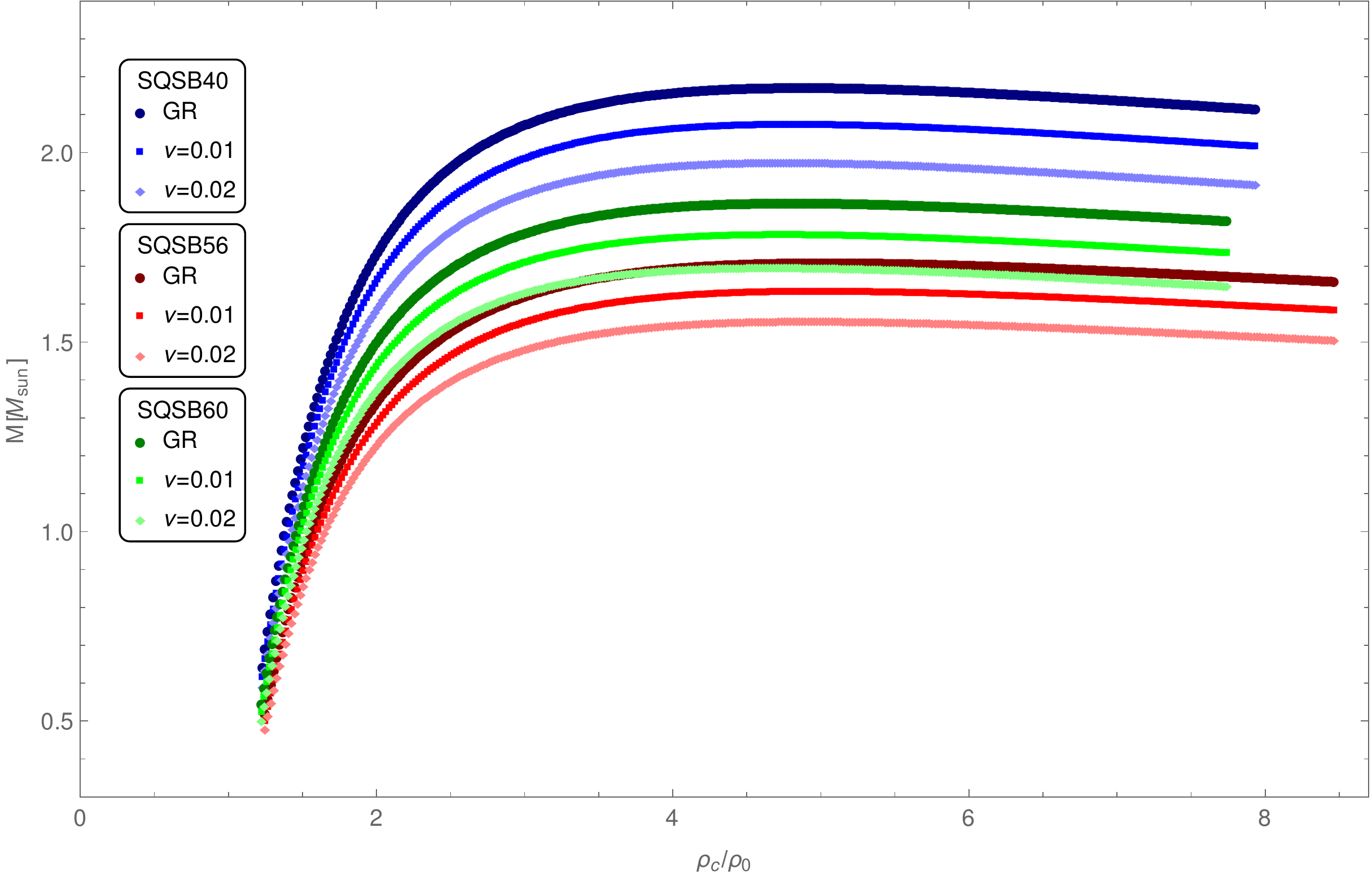} \ 
\includegraphics[width=0.49\textwidth]{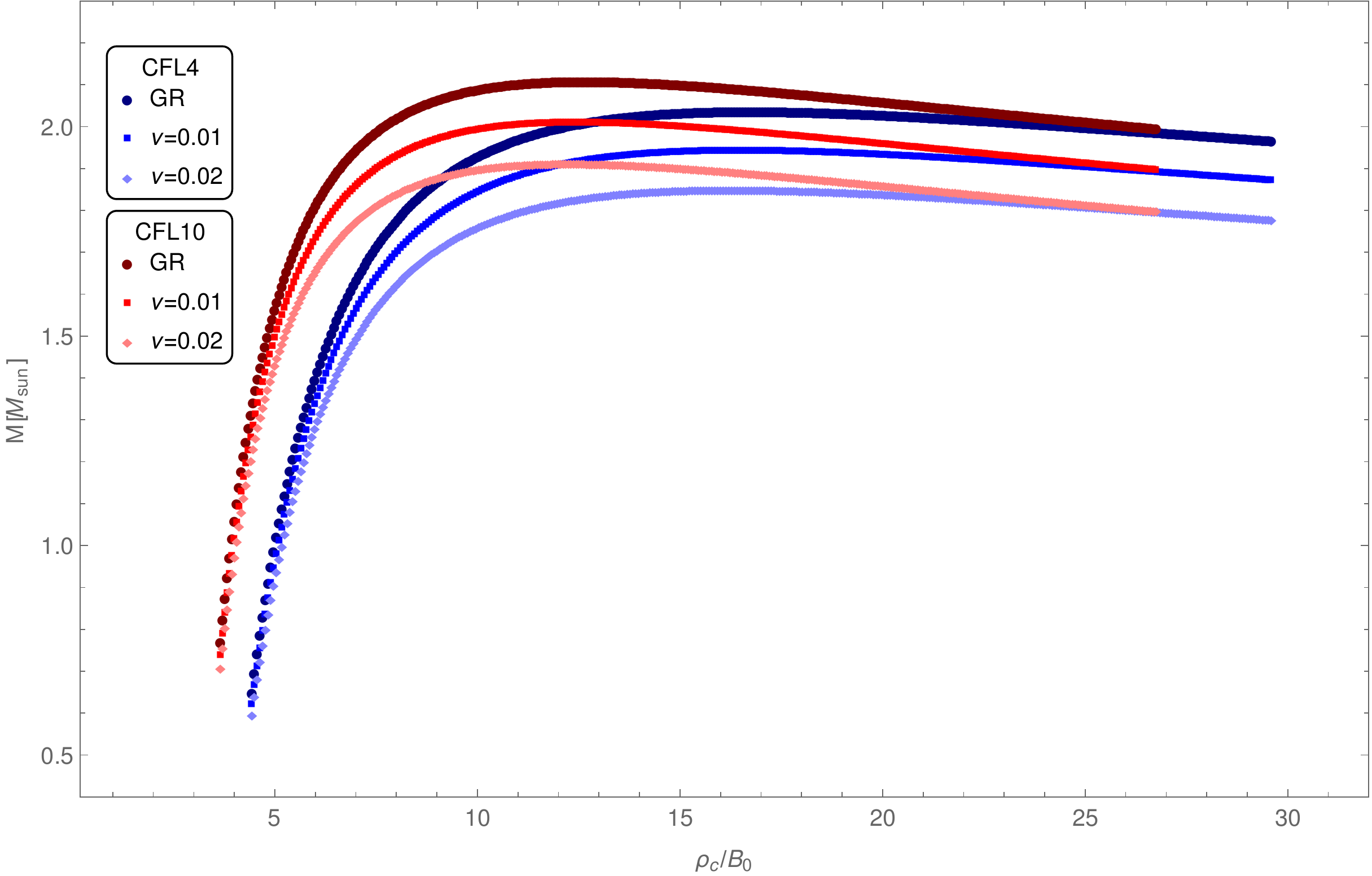}
\caption{
Mass of the star (in solar masses) vs normalized central energy density for GR ($\nu=0$) as well as for $\nu=0.01, 0.02$. 
{\bf Left panel:} Mass vs $\rho_c/\rho_0$ for the 3 MIT bag models, SQSB40 (blue), SQSB56 (red) and SQSB60 (green). 
{\bf Right panel:} Mass vs $\rho_c/B_0$ for the 2 CFL models, CFL4 (blue) and CFL10 (red).
}
\label{fig:6}
\end{figure}

Regarding the stability criteria, we need to verify that $\Gamma > 4/3$~\cite{Moustakidis:2016ndw}, where the adiabatic index $\Gamma$ is defined by
\begin{equation}
\Gamma \equiv c_s^2 \left[ 1 + \frac{\rho}{p} \right],
\end{equation}
as well as that the Harrison-Zeldovich-Novikov criterion~\cite{Harrison,ZN} is satisfied, which states that a stellar model is a stable configuration only if the mass of the star grows with the central energy density, {\it i.e.},
\begin{equation}
\frac{dM}{d \rho_c} > 0\,.
\end{equation}
Fig.~\ref{fig:5} shows that $\Gamma > 4/3$ for the models SQSB56 and CFL10 when $\nu=0.02$, but we have checked that we have obtained almost identical figures for all the cases considered in the present work. The two panels of Fig.~\ref{fig:6} show the mass of the star (in solar masses) versus normalized central energy density both for the MIT bag models and CFL models considered here.

\section{Conclusions}
\label{Sec4}

To summarize our work, in this article we investigated the properties of non-rotating strange quark stars with isotropic matter in Lorentz-violating theories of gravity. To be more precise, we have studied quark stars in Ho\v rava gravity and Einstein-{\ae}ther theory, whose deviations from GR are characterized by a single dimensionless parameter $\nu$. For the quark matter EoS we adopted analytic functions widely used in the literature, both linear and non-linear, corresponding to the simplest version of the MIT bag model and to the color flavor locked state when color superconductivity effects become important at very high densities, respectively. We integrated numerically the generalized structure equations, and we computed the compactness, the radius and the mass of the stars upon matching the interior and the exterior solutions at the surface of the object. We showed graphically the $M_R$ relations for different EoSs as well as for different values of the parameter $\nu$. Our results show that the $M_R$ profiles are shifted downwards as $\nu$ increases, which implies a lower highest mass supported by a given EoS compared to the one obtained in GR for the same EoS. Finally, we have checked that both the energy conditions and stability criteria are fulfilled, and therefore the solutions obtained here are realistic solutions within the framework of the gravitational theories considered in this work.


\vspace{0.3cm}

{\bf Acknowledgments:} We wish to thank the reviewer for useful comments and suggestions.
The authors G.~P. and I.~L. thank the Funda\c c\~ao para a Ci\^encia e Tecnologia (FCT), Portugal, for the financial support to the Center for Astrophysics and Gravitation-CENTRA, Instituto Superior T\'ecnico, Universidade de Lisboa, through the Grant No.~UIDB/FIS/00099/2020. The author D.~V. was supported by Funda\c{c}\~ao para a Ci\^encia e a Tecnologia (FCT) through the Research Grant UID/FIS/04434/2019, and by Projects PTDC/FIS-OUT/29048/2017, COMPETE2020: POCI-01-0145-FEDER-028987 $\&$ FCT: PTDC/FIS-AST/28987/2017, and IF/00852/2015 of FCT. 




\end{document}